\documentclass{ws-ijmpa}
\usepackage[super,compress]{cite}
\begin{document}
\def\b{\bar}
\def\d{\partial}
\def\D{\Delta}
\def\cD{{\cal D}}
\def\cK{{\cal K}}
\def\f{\varphi}
\def\g{\gamma}
\def\G{\Gamma}
\def\l{\lambda}
\def\L{\Lambda}
\def\M{{\Cal M}}
\def\m{\mu}
\def\n{\nu}
\def\p{\psi}
\def\q{\b q}
\def\r{\rho}
\def\t{\tau}
\def\x{\phi}
\def\X{\~\xi}
\def\~{\widetilde}
\def\h{\eta}
\def\bZ{\bar Z}
\def\cY{\bar Y}
\def\bY3{\bar Y_{,3}}
\def\Y3{Y_{,3}}
\def\z{\zeta}
\def\Z{{\b\zeta}}
\def\Y{{\bar Y}}
\def\cZ{{\bar Z}}
\def\`{\dot}
\def\be{\begin{equation}}
\def\ee{\end{equation}}
\def\bea{\begin{eqnarray}}
\def\eea{\end{eqnarray}}
\def\half{\frac{1}{2}}
\def\fn{\footnote}
\def\bh{black hole \ }
\def\cL{{\cal L}}
\def\cH{{\cal H}}
\def\cF{{\cal F}}
\def\cP{{\cal P}}
\def\cM{{\cal M}}
\def\ik{ik}
\def\mn{{\mu\nu}}
\def\a{\alpha}

\markboth{Alexander Burinskii}
{Kerr-Newman electron as spinning soliton}

%
\catchline{}{}{}{}{}
%

\title{KERR-NEWMAN ELECTRON AS SPINNING SOLITON}

\author{ALEXANDER BURINSKII\footnote{}}

\address{Theor.Phys. Lab., NSI, Russian Academy of
Sciences, B. Tulskaya 52 Moscow, 115191 Russia.\footnote{e-mail: bur@ibrae.ac.ru.}\\
first\_author@domain\_name}

\maketitle

\begin{history}
\received{Day Month Year}
\revised{Day Month Year}
\end{history}

\begin{abstract} Measurable parameters of the electron indicate
that its background should be  described by the Kerr-Newman (KN) solution.
 Spin/mass ratio of the electron  is extreme large, and the black hole horizons
 disappear,  opening a topological defect of spacetime -- the Kerr singular ring
 of the Compton size, which may be interpreted as a closed fundamental string to the
low energy string theory. The singular and twosheeted structure of the corresponding
Kerr space has to be regularized, and we consider the old problem of regular source
of the KN solution. As a development of the earlier Keres-Israel-Hamity-L\'opez model,
we describe the model of smooth and regular source forming a gravitating and relativistically rotating soliton based on the chiral field model and the Higgs mechanism of broken symmetry. The model reveals some new remarkable properties:
1) the soliton forms a  relativistically rotating bubble of the
 Compton radius, which is filled by the oscillating Higgs field in pseudo-vacuum state,
 2) boundary of the bubble forms a domain wall which interpolates between the internal  flat background and the external exact Kerr-Newman (KN) solution,
 3) phase transition is provided by a system of the chiral fields,
 4) vector potential of the external the KN solution forms a closed Wilson loop which is quantized, giving rise to quantum spin of the soliton.
5) soliton is bordered by a closed string, which is a part of the general complex stringy structure.

\end{abstract}

\ccode{PACS numbers:}

\section{Introduction and summary}
It is now commonly accepted that black holes (BH) have to be
associated with elementary particles. Physics of black holes is
based on the complex analyticity, which unites them with quantum
and superstring theories and particle physics.

In spite of these evident relationships, Gravity and Quantum
theory are conflicting and cannot be unified in a whole theory.
Similarly,  the path from Superstring theory to particle physics
represents also a still unsolved problem, and as it was  recently
claimed by John Schwarz \cite{Schw}, ``... realistic model of
elementary particles still appears to be a distant dream...'' .

Principal point of the conflict between gravity and quantum theory
is related with the statement on the pointlike and structureless
quantum electron. This point cannot be accepted by gravity, which
requires an extended soliton-like structure of the electron, as a
 field distribution with  a regular energy-momentum tensor in configurational
space. Contrary, quantum theory suggests the claim on the
pointlike structure of the electron, or its statistic description by a wave
function. String theory replaced the
point-like quantum particles by the extended strings and
membrane-like sources. However, the principal quantum particle - electron,
is still considered as point-like. In particular,
 Frank Wilczek  writes in \cite{FWil}: "...There's no evidence that
 electrons have internal structure (and a lot of evidence against it)".
 Similarly, the superstring theorist Leonard Susskind notes
 that  electron radius is "...most probably not much bigger and
not much smaller than the Planck length..",   \cite{LSuss}.
 It should be mentioned that this point of view is supported by
 the high energy scattering, which have not found the electron
structure down to $10^{-16} cm .$ However note,
Quantum Electrodynamic considers an effective  size of a ``dressed electron'',
which corresponds to the Compton region of vacuum polarization. Although space-time
structure of this region is usually not discussed, some hint is
coming from the  relativistic quantum mechanics, which
indicates zitterbewegung of the electron, a lightlike helicoidal motion
following from the Dirac equation.

These partial indications on the peculiar role of the Compton zone
of the electron find unexpectedly strong support from Kerr's gravity.

It was obtained by Carter that the Kerr-Newman (KN) solution, which is exact
solution of the Einstein-Maxwell gravity for a charged and rotating black-hole
(BH), has the gyromagnetic ration $g=2$ as that of the Dirac electron.
Therefore, the four experimentally observable parameters of the electron:
spin $J ,$ mass $m ,$ charge $e$ and magnetic moment $\m $
\emph{indicate unambiguously} that gravitational background of the electron should be
described by the KN solution.
Extremely large spin of the electron with respect to its mass should produce an
over-rotating Kerr geometry without horizon, which displays a naked topological defect
of space-time in the form of the ``Kerr singular ring'' of the radius
$a= \hbar/2m ,$ which is half of the Compton wave length $\lambda_c = \hbar/m .$
This singular ring turns out to be branch line of the space into two sheets resulting
in a two-fold structure of the electron background.
The corresponding gravitational and electromagnetic
fields of the electron are concentrated near the Kerr ring,
forming a sort of a closed string, structure of which turns out to be
close to the described by Sen heterotic string solution \cite{Sen}.
This contradicts to the statements on the structureless electron
and is very far from the its Planck size suggested by superstring
theory. However, it confirms  the peculiar role of the Compton
zone of the "dressed" electron of Quantum electrodynamics,
matches with the known limit of the localization of the Dirac
electron \cite{BjoDr} and indicate relationships with string theory.

There appear two questions:

(A) How does the KN gravity know about one of the principal
parameters of Quantum theory? and

(B) Why does Quantum theory works successfully on the flat
spacetime, ignoring such strong defect of the background geometry?

A small and slowly varying gravitational field could be neglected,
however the stringlike KN singularity forms a branch-line  of the
KS spacetime, and such a topological defect cannot be ignored. A
natural resolution of this trouble could be the assumption that
there is an underlying structure, or even the theory providing the
consistency of quantum theory and gravity.
We conjecture that such underlying structure is to be the Kerr geometry
, complex structure of which indicates close relations to a four-dimensional
version of superstring theory, the ``mysterious" N=2 superstring theory
which is consistent in four dimensions, but does not have the standard
string interpretation \cite{GSW}. It has been suggested in \cite{BurAlter}
that it may be considered as an alternative to the higher dimensional superstring theory.

 In this paper, we consider structure of the
real source of the Kerr geometry. Starting in sec.2 from motivations to consider
the Kerr singular ring as a closed heterotic string, we discuss
  in sec.3 peculiarities of the over-rotating Kerr geometry (without horizons), singular ring, twosheetedness and specifical properties of the Kerr coordinate system. In sec.4 we
  we consider development  the models of source of the KN solution, and in sec.5 consider
   regularization of the KN solution by the Higgs fields, which creates the
 source of the KN solution as a \emph{gravitating soliton} in the
 form of a rotating superconducting disk of the Compton radius
 with a closed relativistic string situating at the disk perimeter. In this section
 we obtain some remarkable peculiarities of the spinning soliton model:

 a) oscillations of the Higgs field with the frequency $\omega =2m ,$ where $m$ is mass of the soliton,

 b) appearance of the quantum loop of the vector potential (Aharonov-Bohm-Wilson loop), which is wrapped around the disk-like source providing quantization of the soliton spin.

In sec.6 may be the most hard for the reading, since we consider there the field aspect of this model, and show that for consistency with gravity we need to extend the Higgs field model to chiral model containing a triplet of the chiral fields. This complication provides a phase transition from the external exact KN solution to a regular source, interior of with is build of pseudovacuum state of the Higgs field resulting in flat metric in the disk-like core of the KN solution.
Therefore we obtain the desirable flat background in vicinity of the KN
source, answering the second above-mentioned question.
Finally, in conclusion we discuss some relations of this model with QED and superstring theory.

\section{Kerr singular ring as a string, first qualitative treatment}

 The Kerr singular ring is generated as a
caustic of Kerr congruence or the focusing line. The KN gravity
indicates  that this string should represent one of the principal
elements of the extended electron structure.

The widespread opinion that the range of interaction for
gravitational field is "tremendously weak and becomes compatible
to other forces only at Planck scale"  is inspired by the usual
analysis of the spherically symmetric Schwarzschild solution,
which has a characteristic radius of the interaction (determined
by the position of horizon, $ r_g=2m $) proportional to the mass
parameter.
 The Kerr geometry breaks this predicate, and moreover, it turns
 this dependence into inverse one, $r_g \sim J/m,$ showing that
 the area of expansion of the Kerr gravitational field is
inversely proportional to mass and proportional to spin of the
system. This unexpected  effect follows from the Kerr relation for
the radius of the Kerr singular ring, $ a = J/m ,$ which shows
that a strong gravitational field may occupy very large region for
the objects of low masses $m$ and large angular momentum $J ,$
which is just the case corresponding to elementary particles. This
seeming paradox has very simple explanation --  gravitational
field of the Kerr solution
 vanishes at the centrum of the solution and concentrates in a
 thin vicinity of the Kerr singular ring,
 forming a type of ``gravitational waveguide'', or string for
 propagation of circular waves, as it was suggested more than forty years
 ago in \cite{IvBur}.
 Although, this simple model of the electron is very naive, it gives
 intuitive explanation to many important facts:
    \begin{romanlist}[(ii)]
 \item  first of all we obtained that the Kerr singular ring represented a waveguide for
 propagation of circular waves, which corresponded to circular motion of a
 massless particle. In the modern terms of the dual string model, there appeared a
 type of the four-dimensional ``compactification without compactification'',

\item mass of the particles originated as energy of the massless
excitations, which is similar to origin of the mass spectrum in
string theory and corresponded to the old Wheeler  model of the
``mass without mass '',

\item the process of mutual transformation of the massive and
massless particles, in particular, annihilation of the electron
positron pair had got natural intuitive explanation,

\item circular motion of the photon with the wave-length $\lambda$
and the energy $E= h c / \lambda $ created relativistic increase
of mass $m=m_0/\sqrt {1+(v/c)^2},$ which followed from the simple
geometric relations,

\item there appeared natural explanation of the ``zitterbewegung'' of the Dirac electron,

\item quantum spin could be interpreted as a consequence of the Bohr quantization of the
 photon waves wrapped around the Kerr ring,

\item the half-integer wave-lengths corresponding to half-integer spin and the spinor
 twosheetedness could be related with twosheeted structure of the Kerr
 space-time,

\item the wave-particle dualism and origin of the de Broglie waves
have got natural explanation.

\end{romanlist}

 It was too much for such a simple model.
Of course, there appeared also problems.
  Principal trouble of this model is the question: What could keep the
  photon   on the orbit of the Compton radius? Estimations of the
  Schwarzschild gravitational field  showed  that it is too weak at
  the Compton  distances. However, contrary to the Schwarzschild solution,
  which has a `range'  of the gravitational field  proportional to
  the mass of the source (radius of the
horizon $r_g = 2m $) \fn{We use the natural units $\hbar=G= c =1 ,$ in which $e^2=\alpha \approx 137^{-1} .$  }, the new dimensional parameter of the KN geometry
  $a=J/m ,$ which grows
with angular momentum $J$ and has the reverse mass-dependence. As
a result, the zone of gravitational interaction determined by
parameter $ a $ increases for the large angular momentum
and small masses, and therefore, it turns out to be essential for
elementary particles. The reason of that is a specific structure
of the KN gravitational field, which concentrates near the Kerr
singular ring,  forming a closed gravitational waveguide -- a type
of the closed gravitational string \cite{IvBur}.

Kerr-Newman (KN) solution has gyromagnetic ratio $g=2,$ as that of
the Dirac electron \cite{Car}, and therefore, at least the
asymptotic gravitational and electromagnetic (em) field of the
electron should correspond to the KN solution with great
precision. Because of that, the charged Kerr-Newman (KN) solution
\cite{KerNew} has paid attention as a classical background of
electron, \cite{Car,DKS,Bur0,Isr,Lop,DirKer,Dym,BurSol,TN,BurQ,BurQ1}.

\section{Over-rotating Kerr geometry and twosheetedness}

The spin/mass  ratio of the elementary particles $ a=J/m$ is
extremely high. In the dimensionless units $c=G=\hbar=1,$ it is
about $ 10^{22} , $ while already for $a/m >1$
 the BH horizons disappear.  It indicates that \emph{spinning
 particles should correspond to over-rotating BH solutions,
 for which the BH horizons disappear, and there appears a naked  Kerr
 singular ring. }
 The electron background acquires a source in the form of a closed
 string of the Compton radius $a = \hbar/ 2m.$

  The over-rotating KN has simple representation in the  Kerr-Schild (KS) form
  of metric
 \cite{DKS},
 \be g_\mn=\eta _\mn + 2H k_\m k_\n \label{ksH} , \ee where
$\eta _\mn$ is metric of an auxiliary Minkowski background in
Cartesian coordinates ${\rm x}= x^\m =(t,x,y,z),$
 $H=H(x)$ is a scalar function, and  $k^\m =(1, \vec k)$ is a null 4-vector
 field. Using signature $(-+++)$ we obtain $k_\m k^\m =(\vec k)^2 - 1 =0 ,$
 and consequently $\vec k$ is a unit spacelike vector field, $(\vec k)^2=1 .$
The use of auxiliary Minkowski background allows to avoid
dependence of the metric on position of the horizon. The KS metric
form  is extreme simple, and one wonders how it can describe the
Kerr metric which is known as very complicate.
 The reason is hidden in the very complicate form of the vector
 field $\vec k ,$ which represents a vortex of the so-called Pricipal Null
 Congruence (PNC), or simple Kerr congruence. The form of field $\vec k
 (x)$is shown in Fig.1.

 \begin{figure}[ht]
\centerline{\epsfig{figure=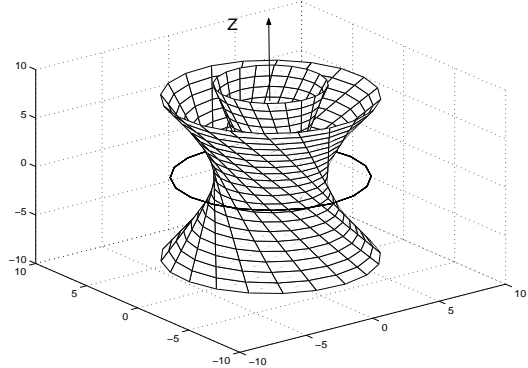,height=5cm,width=7cm}}
\caption{ Twistor null lines of the Kerr congruence are focusing
on the Kerr singular ring, which forms a branch line of space in
two sheets.}
\end{figure}

Four vector $k_\m (\rm x),$ tangent direction to PNC,  represents
a family of the lightlike lines, which  form a skeleton of the
Kerr geometry. These lines are twistors. The reader should not be
frightened by word `twistor'. Indeed, twistor theory is
excessively mathematized, but physical meaning of a twistor is
very simple: it is a  geodesic line of a
photon (lightlike or null line) passing aside of the coordinate origin
or position of the observer. Twistorial Kerr congruence
means that each point of the Kerr spacetime is polarized, and has
a selected lightlike direction. All the tensor fields of the Kerr
geometry turns out to be aligned with this selected  lightlike
direction.

The factor $H$ in (\ref{ksH}) is a scalar function, which for the
charged Kerr-Newman (KN) solution takes the form  \be H = \frac
{mr-e^2/2} {r^2 + a^2 \cos^2 \theta},  \label{H}\ee where $r=0$
and $\theta =0$ are oblate spheroidal coordinates which are
adapted to twisted and twosheeted structure of the Kerr
congruence. The KS formalism uses a few different coordinate
systems, which allows one to adapt treatment to different aspects
of the Kerr solution and simplify calculations. The oblate
spheroidal coordinates represent a family of oblate ellipsoids
$r=const.$ and confocal family of the hyperboloids
$\theta=const.,$ see Fig.2. The Kerr oblate spheroidal coordinates
$r$ and $\theta$ and $\phi_K$ are related with minkowskian
coordinates as follows, \be x+iy  = (r + ia) e^{i\phi_K} \sin
\theta, \qquad z = r\cos\theta, \qquad \rho = t - r ,
\label{oblate} \ee where $\rho$ is the additional `light cone'or
`retarded time' coordinate, which describes propagations of the
waves along the null rays of the Kerr congruence.  However, it
will not be essential for our treatment here.

\begin{figure}[h]
\centerline{\epsfig{figure=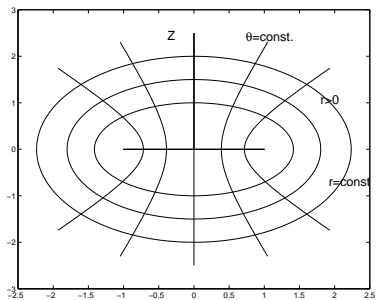,height=4cm,width=5cm}}
\caption{\label{label}Oblate coordinate system $r, \ \theta $ with
focal points at $r=\cos\theta = 0$ covers the space-time twice,
for $r>0$ and $r<0.$ }
\end{figure}

The naked Kerr singular ring which corresponds to $r=0$ and $
\theta =\pi/2.$ It is a focus line of the null directions $k^\m ,$
and simultaneously, it is  a branch line of the Kerr space into
two sheets, $r <0 ,$ and  $r>0 .$ The oblate spheroidal
coordinates are twosheeted and adapted to two sheets of the Kerr
geometry. One sees that if congruence is in-going on the sheet
$r<0 ,$ its analytic extension is to be out-going on the sheet
$r>0 .$  The null directions $k_\m ,$ and metric (\ref{ksH}) turn
out to be different on the `positive' and `negative' sheets.
 We obtain that the  background is spoiled at the Compton
 distances $ a=\hbar/2m ,$ getting a strong topological defect
 in the form of the Kerr singular ring and twosheetedness.
 This twosheeted space is reminiscent of the old Einstein-Rosen bridge,
 or the discussed later Wheeler's worm-hole, which allows one to create a
 `charge without charge'  preventing the space-time from formation of the
 singularities.
Strong curvature of the background in the Compton region indicates
that a \emph{regularization, analogous to regularization of
charge,  has to be executed for  the metric too}. In fact, this
strong breakdown of space indicates  that the Compton region has
to be a zone of new physics. In accord to results of Quantum
Electrodynamics this zone should be related with processes of
vacuum polarization.

\section{The `bubble' and string-like sources of KN solution.}

\textbf{Bubble source.} It is remarkable, that the Kerr geometry
gives unambiguous answer on the size and shape of this zone,
\cite{Lop}. One sees that the KS metric (\ref{ksH}) becomes flat
for the vanishing function $H ,$ and therefore, one should set
$H=0$ for regularization of the metric in the source of KN
solution. The boundary of this source is determined from (\ref{H})
by the value of the Kerr ellipsoidal radial coordinate \be r = r_e
= e^2/(2m)  \label{re} ,\ee inside of which, $r<r_e ,$ the
space-time should be set as flat, $H=0.$ We obtain that the source
of KN solution should represent an ellipsoidal shell with flat
interior which resembles a \emph{vacuum bubble.} The bubble has
form on a highly oblate disk. One sees from the coordinate
relations (\ref{oblate}), that for $r=r_e ,$ radius of the disk
(corresponding to  $z=0$) is $ \sqrt{x^2 +y^2} =\sqrt{r_e^2 + a^2}
,$ while the disk thickness (corresponding to $x^2 +y^2 =0$) is
determined by $z_{disk} = r_e ,$ and therefore, the ratio
thickness/radius turns out to be $r_e/\sqrt{r_e^2 + a^2}.$ In the
natural units ($\hbar=G=c=1$) $e^2 \approx 137^{-1},$ and $r_e
=e^2/(2m) \approx 137^{-1}/(2m).$ From the Kerr relation
$J=ma=\hbar/2=1/2 $ , we have $a= 1/(2m),$ and therefore $r_e =
a/137 << a .$ We obtain that the ratio thickness/radius is close
to $r_e/a = 137^{-1},$ and determined by the fine structure
constant $\alpha =137^{-1} .$ The fine structure constant acquires in
Kerr's electron a geometric meaning!
 A very important specification was given by Hamity \cite{Ham},
 who noticed that the disk should be rigidly rotating reaching the
 velocity of the light  at the edge border.

Let's look now at the electromagnetic (EM) field. The vector
potential of the KN solution is given by \be A^\m_{KN} = Re \frac
e {r+ia \cos \theta} k^\m  \label{ksA}. \ee One sees that it is
also proportional to the null direction $k^\m ,$ and therefore, it
is also aligned with the Kerr congruence (PNC), i.e. \be A^\m_{KN}
g_\mn k^\n =0 . \ee This is principal property of the
algebraically special solutions, that all the tensor quantities
are aligned with PNC, which simplifies solutions of the field
equations.

  This source was suggested by L\'opez as a classical model of an
extended electron in general relativity \cite{Lop}. Like the other
shell-like models, the L\'opez model was not able to explain the
origin of Poincar\'e stress, and the necessary tangential stress
was introduced by him phenomenologically, as a distribution over
the surface of the rotating shell.

Meanwhile, the necessary tangential stress appears naturally in
the field models of domain walls. The corresponding field model of
the domain wall bubble was suggested in \cite{BurBag,BEHM} and
developed in \cite{BurSol} as a gravitating soliton model, in
which the Higgs field is concentrated inside the bubble in a
pseudo-vacuum superconducting state and performs regularization of the KN
solution.

The Higgs field pushes the em field from the bubble. As a result,
 the em field is regularized, acquiring the cut-off parameter
 $r_e $ leading to maximal value of the vector potential
\be A^\m_{max} = Re \frac e {r_e} k^\m = Re \frac {2m} e k^\m
\label{Amax} ,\ee which is reached on the boundary of the disk
corresponding to $\cos \theta=0 .$

The regularized KN space tends to flat near the source, in
agreement with the requirements of Quantum theory. However, it
should be noted, that performing regularization we distorted the
original KN solution. Submission of the regular EM field and
metric in the Einstein-Maxwell system of the field equations will
produce the additional charge,  currents and matter on the surface
of the bubble in the form of a density distributions described by
$\delta$-function \cite{Lop}. These tensor densities take a simple
diagonal form in a corotating system of coordinates, which
evidences that the disk-like bubble has to be rigidly rotating,
and the linear velocity at the boundary of the disk is to be close
to the velocity of light. In the recent development, this model
takes the form of a gravitating soliton \cite{BurSol}, in which
the infinite thin shell of this bubble is replaced by a domain
wall boundary, and the bubble is not empty, but is filled by the
Higgs field in a pseudovacuum state. A special set of the chiral
fields performs a phase transition from the external exact KN
solution to the Higgs field inside the bubble providing the flat
internal metric and pseudovacuum state of the Higgs field.

\textbf{String-like source.} Let's now consider stringy
interpretation of the Kerr singular ring, which was initially
considered  as an alternative model of the source of Kerr
geometry. It was suggested in \cite{IvBur} to consider the Kerr
ring as a closed relativistic string similar to strings of the
dual models. The massless equations of the relativistic strings
create the massive states from energy of string excitations. This
mechanism is similar to the Wheeler's idea of `geon',
gravitational-electromagnetic object with `mass without mass'.
Mass of the `geon' is generated by energy of the electromagnetic
field,
 in particular, by photons traveling on the circular orbits.
 It was suggested in \cite{IvBur} that the Kerr singular ring may
 represent a type of gravitational waveguide which keeps the electromagnetic (em)
 or spinor waves in the orbital motion, \cite{Bur0,IvBur,BurQ}.
 Twenty years later, the string-like solitonic solutions with traveling waves were
 considered  as fundamental strings solutions to low energy string theory
 \cite{Garf,DGHW}.
 It was been shown in \cite{BurSen} that the field structure of the Kerr singular ring
is similar to the structure of the fundamental  heterotic string
in  the obtained by Sen solutions to low-energy string theory \cite{KerSen,Sen}.
Later on, it was obtained that this string is only "a tip of the iceberg",
and a wonderful stringy system related with N=2 superstring theory is
indeed hidden in the complex Kerr geometry \cite{BurAlter}.

\section{Regularization of the em field by the Higgs fields.}

\subsection{Basic field equations.} Regularization of the em
field in the bubble-source model is
performed by the Higgs mechanism of broken symmetry, which was
used in many particle-like models, like the t`Hooft-Polyakov
models and the the Nielsen-Olesen \cite{NO} field model for the
string-like solution in superconductivity.

The em field in vacuum and in the Einstein-Maxwell theory is
massless and long-distant. Presence of the Higgs field gives a
mass to the em field making it short-distant. The em field cannot
deeply penetrate in the regions occupied by the Higgs field. The
depth of penetration $\delta \sim 1/m$ depends on the acquired
mass $m .$ Because of that the Higgs field is used for description
of the Meissner effect, interaction of the em field with ideal
conductors and superconductors. There is also an opposite
influence: the strong em field expels the Higgs field. For
example, it penetrates in a superconductor in the form of vortex
filaments, as it was described by  Abrikosov solutions. The Higgs
field model was used in the t`Hooft and Polyakov particle-like
models of the magnetic monopole, and also was used by Nielsen and
Olesen as a model of the dual relativistic string in a
superconducting media.  In particular,
the Lagrangian used in \cite{NO}  for the complex Higgs field $\Phi (x)$
interacting with the em vector field  $A^\m$ has the form  \be
{\cal L}_{NO}= -\frac 14 F_\mn F^\mn + \frac 12 (\cD_\m
\Phi)(\cD^\m \Phi)^* + V(|\Phi|), \label{LNO}\ee where $ \cD_\m =
\nabla_\m +ie A_\m $  are to be covariant derivatives, and $F_\mn
= A_{\m,\n} - A_{\n,\m} .$
We can use the similar field model.  In the vicinity of the KN source we have $H\approx 0$ leading to flat space-time, which allows us to consider $\nabla_\m$ as
 flat derivatives and to set $\nabla _\n \nabla^\n = \Box .$
 This Lagrangian gives rise to the system of equations for the coupled
 Maxwell-Higgs system
\bea \cD^{(1)}_\n \cD^{(1) \n} \Phi &=& \d_{\Phi^*} V  , \label{PhiIn} \\
\Box A_\m = I_\m &=&  e |\Phi|^2
(\chi,_\m + e A_\m). \label{AIn} \eea

This system of the coupled equations describes an interplay of the  Higgs and Maxwell field separated by some contact boundary and their mutual penetration through this boundary.
Exact solutions of this system are known only for some particular cases of the  potential $V$ and only for flat boundary.  Analysis of the solutions requires usually numerical calculations or strong simplifications.
 It should also be noticed that the most of the known solutions to the Maxwell-Higgs system describe the localized em field confined in a restricted region of space surrounded by the Higgs field, i.e. a cloud of the gauge massless fields surrounded by a superconducting media formed by Higgs field.  Meanwhile, our task is to consider opposite situation, in which the Higgs field is localized inside the bubble and surrounded by  the massless em field extended to infinity. At first sight, this difference seems inessential, but indeed, our case of  interest becomes much more complicated and requires introduction of the several Higgs-like fields and the \emph{potential} $V$ of an especial domain wall form. We will discuss resolution of this problem in the final section of our treatment, but now we  consider some extra simplifications, which will allow us to clarify basic peculiarities of the regular source of the KN solution.

\subsection{The source without rotation.} We can further simplify
the problem considering the source without rotation. By setting $a
=0 ,$ we obtain that the Kerr singular ring shrinks to a point and
the vector potential (\ref{ksA}) takes the spherically symmetric
form \be A^\m_{0} = \frac e r k^\m \label{ks0}, \ee where $r$ is
the usual real radial coordinate, $r=\sqrt{x^2 +y^2 +z^2} .$ The
null vector field $k^\m$ turns into a spherically symmetric system
of the four-vectors $k^\m =(1, \vec n),$ where $\vec n = x^\m/r $
represents a hedgehog of the unit radial directions. The metric
(\ref{KSh}) and vector potential (\ref{ks0}) correspond to the
Reissner-Nordstr\"om solution in the Kerr-Schild form.

The bubble-source filled by Higgs field takes in this case the
spherical form, and the maximal value of regularized vector
potential will be again \be A^\m_{max} = Re \frac e {r_e}
k^\m = Re \frac {2m} {e} k^\m\label{Amax0} ,\ee which corresponds to the
classical model of the electron as a charged sphere with a unique
difference that the sphere is replaced by a superconducting
ball, interior of which is filled by a complex Higgs field
$\Phi(x)= \Phi_0 \exp \{i\chi (x) \} ,$ where $\Phi_0$ is the vacuum
vacuum expectation value (`vev') of the Higgs field.

Therefore, we replace the equation (\ref{PhiIn}) by assumption on
the sharp boundary of the bubble, setting $\Phi(x)=0$ outside the
bubble and  $\Phi(x)=\Phi_0 ,$ for $r<r_e .$ The corresponding em
field is massless and has the form (\ref{ksA}) for $r> r_e ,$,
while penetrating inside the bubble it should satisfy the eq.

\be \Box A_\m =I_\m = e |\Phi|^2 (\chi,_\m + e A_\m) \label{Main}.\ee

If the phase $\chi$ is a constant, this equation turns into $ \Box A_\m = m_v^2  A_\m ,$ where
$ m_v=| e \Phi| $ is the mass acquired by vector field $A_\m $ due to Higgs mechanism.
Here we have another action of the Higgs field, which is related with compensating role of the phase $\chi$ representing a space-time function $\chi (x)$.
Splitting $A_\m = (A_0, \vec A) $ into timelike component $A_0 =\frac e r $ and the
radial field $\vec A = \frac e r \vec n \equiv e d \ln r ,$ one can drop \emph{formally}
the radial part as a full differential, which will not products the strengths of
the em field $F_\mn .$
In accord to (\ref{Main}), the time-like component $A_0=\frac e r $ could create the charge
\be \rho= I_0 = e |\Phi|^2 (\chi,_0 + e A_0) \label{I0} \ee
However, the charge and current have to be expelled from superconducting interior of the bubble to its boundary. It sets the condition  $I_0=0 $ for $r<r_e $ resulting in the
the relation $ \chi,_0 + e A_0^{in} =0 ,$ which fixes the value of component $A_0$ inside the bubble as a constant determined by the frequency  of oscillations of the Higgs field $\omega ,$ \be  A_0^{(in)} = - \frac 1e \d_t \chi = - \frac 1e \omega .\label{AOinomega} \ee
 Note, that
the left side of the (\ref{Main}), $\Box A_0^{(in)}=0 ,$ is
satisfied by the constant value of $A_0^{(in)}.$ At the boundary
of the bubble $r_e ,$ the constant value $A_0^{(in)}$ has to be
matched with external solution $A_0 =\frac e r ,$ \be A_0^{(in)} =
- \omega /e = \frac e {r_e} .\ee Using $r_e =e^2/(2m) ,$ we arrive
at very important result that \emph{the Higgs field forms a
coherent vacuum state oscillating with the frequency } \be
|\omega| = 2m . \label{omega} \ee Solitonic solutions of this type
were called ``oscillons''. Solutions of this type were first considered
 by G.Rosen \cite{GRos} and Coleman \cite{Coleman} (Q-ball). Examples of the
 oscillon solutions are spinning Q-balls \cite{VolkWohn,Grah} and the
 bosonic star solutions \cite{BosStar}.

\subsection{Inclusion of the rotation.}

Let us write now the Higgs phase in the form  $\chi =\omega t +
n\phi + \chi_1(r),$ where we have taken into account the constant
frequency $\omega ,$ periodicity in $\phi$ and some dependence on
$r.$ The Higgs field inside the bubble can be represented as
follows \be \Phi = |\Phi| \exp \{ i(\omega t + n\phi + \chi_1(r))
\} ,\ee where the azimuthal coordinate $\phi$ my be expressed via
cartesian coordinates as follows $\phi=-i\ln[(x+iy)/\rho], \
\rho=(x^2+y^2)^{1/2} .$

In the rotating case the KN gauge field $A_\m ,$
given by (\ref{ksA}), is twisted, since it is aligned with
 tangent vector to twisted Kerr congruence, $k_\mu$.
 One sees, that the basic expressions for the Kerr
metric (\ref{ksH}) and the em field (\ref{ksA}) are extreme
simple. The complicated structure of the Kerr solution is
concentrated in the form of vector field $k^\m(x)$ which may be
described in differential form in the Cartesian or the Kerr
angular coordinates. The Cartesian representation is more
important from theoretical point of view, since it shows  relation
to twistors, the Kerr theorem and superstrings.\cite{BurAlter} For
our aims here, it is enough to give $k^\m$  in differential form
expressed in the Kerr angular coordinates \cite{DKS} \be k_\m
dx^\m = dr - dt - a \sin ^2 \theta d\phi_K . \label{km} \ee

The Kerr azimuthal coordinate $\phi_K$  has a very specific form
determined by the relation (\ref{oblate}). It is inconsistent with
the standard angular coordinate $\phi$ of the Higgs field, and
their differentials are related as follows \be d\phi_K =d\phi
+\frac {adr}{r^2+a^2} \label{dphidphiK}.\ee Fig.4 shows that the
spacelike part of $k^\m = (1, \vec k),$ $\vec k$ is tangent to
Kerr ring in the equatorial plane at $r=0.$ It means that the Kerr
ring is lightlike, i.e. it slides along itself with the speed of
light.

\begin{figure}[h]
\centerline{\epsfig{figure=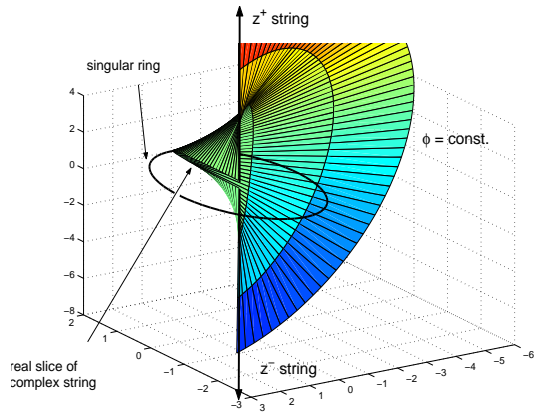,height=5.5cm,width=7cm}}
\caption{\label{label} The Kerr surface $\phi=const.$ The Kerr
congruence is tangent to singular ring at $\theta=\pi /2 .$}
\end{figure}

The KN vector potential (\ref{ksA}) takes the form \be A_\m dx^\m
= \frac {-er}{r^2+a^2\cos^2 \theta} [dr - dt - a \sin ^2 \theta
d\phi_K] .\ee Using (\ref{dphidphiK}), one can express it
 in terms of $d\phi $ and obtain
\be A_\m dx^\m = \frac {-er}{r^2+a^2\cos^2 \theta} [dt + a \sin ^2
\theta d\phi ] + \frac {e r dr} {(r^2 +a^2)} ,\label{Aphiin} \ee
which shows that radial component $A_r$ represents a full
differential and can be dropped.\fn{Inside the bubble $A_r$ is
compensated by the $\chi_1(r)$  component of the Higgs phase.} At
the external side of the bubble boundary, $r=r_e +0 = e^2/2m ,$
the the potential takes the value $ A_\m(r_e) dx^\m = \frac{-e
r_e} {r^2_e +a^2 \cos^2 \theta}[dt + a \sin ^2 \theta d\phi ] , $
and in the equatorial
 plane, $\cos\theta=0,$  it forms the loop
 wrapped along the $\phi$-direction, \be A_\m (r_e) dx^\m
= - \frac{2m} {e}[dt + a d\phi ], \ee  with tangent component
$A_\phi (r_e)= - 2ma/e .$ This field is penetrated inside the
bubble, \be A_\m^{(in)} dx^\m = - \frac{2m} {e}[dt + a d\phi ],
\label{Ain} .\ee However, in agreement with (\ref{Main}), it is
compensated by the phase of Higgs field, resulting in cancelling
of the corresponding circular current inside the bubble, \be
I_\phi^{(in)}=0 \ \Rightarrow \chi,_\phi = - e A_\phi^{(in)} .\ee
It implies periodic dependence of the Higgs field on $\phi$, $
\Phi \sim \exp \{i n \phi \} ,$ with integer $n.$ There appears
the closed Wilson loop of the potential along the rim of disk-like
source. The integral over the loop \be S=\oint eA_\phi (r_e)
d\phi=-4\pi ma \label{WL} \ee has to be matched with periodic
incursion of the Higgs phase $2\pi n,$ and therefore, it turns out
to  be quantized. Using the KN relation $J=ma ,$ we obtain
 wonderful result that  the quantum Wilson loop of the KN em
field $ e \oint A^{(str)}_\phi d\phi=-4\pi ma \label{WL} $
requires \emph{\textbf{quantization of the total angular momentum
of the soliton, $J=ma=n/2, \ n=1,2,3,...$}}

Note, that this result together with the result related with Eq. (17)
follow unambiguously from the form of KN solution and condition (\ref{re}), and therefore, these results follow only from the requirement to get the KN source with a flat internal metric -- from the \emph{requirement of consistency} with flat background of quantum theory.
 This property was never observed in the other spinning Q-balls and the bosonic star
solutions.

\emph{Extra note.} By construction of the solution, the time-like
and $\phi$ components of
 the obtained vector field are continuous at  the boundary of the
 bubble $r_e.$ However, there is discontinuity in their radial derivatives,
 which has to generate circular currents on the bubble boundary. Practically,
 the boundary is not sharp, and the vector field should have a finite depth
 of penetration $\delta_v ,$  and derivatives should be smoothed by this
skin effect. The real values of the vector potential   will differ
from the obtained background solution (\ref{Ain}) in the boundary
layer $ r \in [r_e , \quad r_e-\delta_r].$
 Denoting this deviation as $ \delta A_\m^{(bound)} = A_\m - A_\m^{(in)} $ in (\ref{Main})
 we obtain the equation
 \be \Box \delta A_\m^{(bound)} =I_\m = e^2 |\Phi|^2 \delta A_\m^{(bound)} \label{skin}\ee
 which shows that the em field  acquires via Higgs mechanism the
 mass $m_v=e |\Phi|,$ which produce the charge and ring-like current with  the depth
 of penetration $\delta r \sim 1/m_v .$
It may be interpreted as a string-like massive and charged vector
meson residing at the circular boundary of the bubble.

\section{Full action: potential, domain wall, gravity and phase transition.}

\subsection{Phase transition}
We have to consider now the role of the potential $V(\Phi)$ in
formation of the boundary condition and in the phase transition
from the external vacuum state corresponding to KN solution
to some false-vacuum flat state inside the
bubble source. In this point our model differs essentially
from the other solitonic models, as well as from the known
oscillon and Q-ball models. The renormalizable quartic potential
$V= (|Phi|^2 -v )^2 ,$ is generally used  for  the Higgs field
models with broken symmetry.  In particular, is was used by Graham
for the oscillon model in \cite{Grah} and in the well known MIT
and SLAC bag models. This is inappropriate for the gravitating
soliton models, since presence of the Higgs field outside the source gives a mass
to the external gravitational and EM fields turning them in the short-range fields.
In particular, in our case it should distort the external KN field.
Therefore, we have to use a different scheme of phase transition, in which the external vacuum state would be unbroken, and consequently, the Higgs
field should be concentrated inside the source. This turns out to be a nontrivial task which
cannot be solved by the typical quartic potential.

The similar problem appears in the Vilenkin-Witten model of the
superconducting cosmic string \cite{Wit}, and to resolve this
problem Witten used   the $U(1)\times \tilde U(1) $ field model
\cite{Wit} which contains two Higgs field: one of which $\Phi^1$
is concentrated inside the superconducting source, while another
one $\Phi^2$ takes the complementary domain extended up to
infinity. These two Higgs field are charged and adjoined to two
different gauge fields $A^1$ and $A^2 ,$ so that when one of them
is long-distant in some region $\Omega$, the second one is
short-distant in this region and vice-verse. This model is
suitable for our case, however, in \cite{BurSol} we used a
 supersymmetric scheme of phase transition \cite{WesBag} suggested by Morris
 \cite{Mor}. The
Morris potential $V$ depends on two charged Higgs-like complex
fields $\Phi$ and $\Sigma $ and one auxiliary uncharged real field
$Z ,$ which are combined in the superpotential
 \be W= \lambda Z(\Sigma \bar \Sigma -\eta^2) +
(Z+ \m) \Phi \bar \Phi , \label{WMor}\ee where $\ \m, \ \eta, \
\lambda$ are real constants.   In accord to theory of the chiral
superfields \cite{WesBag}, the potential $V$ is determined from
the superpotential $W$ by the relation \be V(r)=\sum _i |\d_i W|^2
, \label{VW}\ee where $ \d_1 = \d_\Phi , \ \d_2 = \d_Z , \ \d_3 =
\d_\Sigma ,$ and $\Phi^{(i)}, \ i=1,2,3$ forms triplet of the
chiral fields\fn{(\ref{VW}) assumes that $\Phi^{(i)}$ and
$\bar\Phi^{(i)}$ are independent fields.} \be \Phi^{(i)} = \{\Phi,
Z, \Sigma \}. \ee

Vacuum states $V_{(vac)} =0$ are determined by the conditions
$\d_i W =0 .$ It is easy to obtain for (\ref{WMor}) two solutions:

(I) vacuum state:   $ \ Z=- \m; \ \Sigma=0; \ |\Phi|=
\eta\sqrt{\lambda},$ corresponds to $W_I=\lambda\m \eta^2  ,$

(II) vacuum state:  $ \ Z=0; \ \Phi=0; \ \Sigma=\eta ,$
corresponding to $W_{II}=0 .$

One can identify the field $\Phi$ as the main Higgs field, and the
state (I), where $|\Phi|>0,$ as the false-vacuum state of the
Higgs field inside the bubble. Then the state (II) ,  where
$|\Phi|=0 ,$  will be identified with external vacuum state with
the non-zero Higgs-like  field $|\Sigma| = \eta^2 > 0 .$

The coupled with gravity action reads

\be S = \int \sqrt{-g} R d^4 x(\frac R {16 \pi G} + {\cal
L}^{mat}) ,\ee

where the full matter Lagrangian takes the form

 \be {\cal L}^{mat}= -\frac 14 F_\mn F^\mn + \frac 12
\sum_i(\cD^{(i)}_\m \Phi^{(i)})(\cD^{(i) \m} \Phi^{(i)})^* + V
\label{L3} ,\ee which contains contribution from triplet of the
chiral field $\Phi^{(i)}.$

The potential $V$ (\ref{VW})  is positive by definition (\ref{VW})
and forms a domain wall interpolating between the internal
false-vacuum state (I) ($V_{(int)}=0$) and external
gravi-electro-vacuum state (II) ($V_{(ext)}=0 $) corresponding to
KN solutions of the Einstein-Maxwell field equations.

The covariant derivations $\cD^{(i)}_\m =\nabla_\m +ie A^{(i)}_\m
$ contain in general case three different gauge fields $A^{(i)}.$
However, in our case we need only one gauge field $ A_\m $
associated with the principal chiral field $\Phi^{(1)} \equiv \Phi
.$ Therefore, we set  $F_\mn = A_{\m,\n} - A_{\n,\m} $ and the
corresponding covariant derivations will be $\cD^{(1)}_\m
=\nabla_\m +ie A_\m , \ \cD^{(2)}_\m =\cD^{(3)}_\m =\nabla_\m .$
the field $\Phi^{(2)} \equiv Z$ is uncharged and $A^{(2)} =0 .$

It should be noted that generalization to triplet of the chiral
field may be useful for generalization of this model to
Salam-Weinberg theory.\fn{This idea is suggested by Th.M.
Nieuwenhuisen.}

\subsection{Problem of the exact solutions.}
We have now to turn to Einstein equations

\be R_\mn -\frac 12 g_\mn = 8\pi G T^{(mat)}_\mn \label{Eeqs}\ee

\noindent {\it The stress-energy tensor} may be decomposed into pure em part
and contributions from the chiral fields  \bea T^{(mat)}_\mn
= T^{(em)}_\mn + \label{T3} \\
\nonumber   \delta_{i\bar j}(\cD^{(i)}_\m \Phi^i)\overline
{(\cD^{(j)}_\n \Phi^j)} &-& \frac 12 g_\mn[\delta_{i\bar
j}(\cD^{(i)}_\lambda \Phi^i)\overline {(\cD^{(j)\lambda} \Phi^j)}
+V] , \label{Tmat}\eea
and we have to consider three different regions: external zone, $r\ge r_e,$  transition zone $(r_e -\delta _ r) \le r_e,$ and zone of flat interior $ r < (r_e -\delta _ r).$

 {\it  In the external zone, $r\ge r_e,$}  we have
 $ V^{ext}=0 .$ The unique nonzero chiral field $\Sigma$ is constant,
 and therefore, all the derivatives
$\cD^{(i)}_\m \Phi^{(i)}$  vanish. As a result  $T^{(tot)}_\mn$ is
reduced to $ T^{(em)}_\mn ,$ and we obtain the
Einstein-Maxwell field equations are satisfied and for the external KN
electromagnetic field they result in the external KN solution.

{\it For interior of the bubble, $ r < (r_e -\delta _ r)$}
we  have also $V^{int}=0 ,$ and the unique nonzero Higgs field is
$ \Phi(x) = |\Phi(x)|e^{i\chi(x)} .$

The Lagrangian (\ref{L3}) is reduced to (\ref{LNO})
with $V(r)=V^{int}=0 ,$ which leads to eqs.
\bea \cD^{(1)}_\n \cD^{(1)\n} \Phi &=& 0 ,  \label{cD1}\\
  \nabla _\n \nabla^\n A_\m &=& I_\m = \frac
12 e |\Phi|^2 (\chi,_\m + e A_\m). \label{APhiIn} \eea

 The only variable chiral field in the flat interior is the oscillating Higgs field,
 and we have to consider it in more details. One sees that the term
\be \cD^{(1)}_t \Phi= (\d_t +ie A^{(in)}_0) |\Phi|\exp \{i\omega t \} =
i (\omega + A^{(in)}_0) \Phi \ee  is cancelled in agrement with (\ref{AOinomega}),
and therefore, (\ref{cD1}) is satisfied.

For flat  interior the second eq. (\ref{APhiIn})  reduces to
the system (\ref{Main}) and we obtain all the consequences considered
in  sec.5. Therefore,  $T^{(mat)}_\mn =0$  and the
Einstein-Maxwell equations are trivially satisfied for flat interior.

One sees that considered
in sec.5 em solutions together with the discussed here Einstein-Maxwell system,
are consistent with the sharp boundary between the external and internal regions,
and the limit $\delta_r \to 0 ,$ may be interpreted as a\emph{ thin wall approximation}.

In this limit, external KN metric  matches continuously with
flat interior of the bubble and turns out to be consistent with
the stress-energy tensor of the external KN solution and flat
interior of the bubble. However, in the thin wall  limit, there appears
discontinuity in the first derivatives of the metric.
Because the Einstein equations contain the second
derivatives of the metric, the stress-energy tensor has a
$\delta-$ function singularity at the thin wall \cite{CvGrifSol}. In
this case of thin wall, the analysis was usually performed  on the basis of
Israel's formalism of singular layers \cite{Isr,CvGrifSol}, or in
terms of generalized functions of  the  theory of distributions,
\cite{Lop}.  The obtained solution may be considered as
a consistent with gravity thin wall approximation.
However, internal structure of the wall is indeed a non-trivial
and  very important  problem.

\subsection{Beyond the thin wall approximation.}

Beyond the thin wall approximation, we have to consider as the third zone
the zone of phase transition $ (r_e -\delta _ r) <r < r_e .$

\emph{Zone of phase transition} is the practically inaccessible for analytic
solutions. Up to our knowledge, analytic solutions of the \emph{similar}
problems with the Higgs field are unknown even for the simple spherical configuration.
The known  analytic solutions obtained for the
vacuum domain walls with planar geometry have a kink-like form,
and the typical stress-energy tensor for domain wall has the form
 \be T^\m_\n ={\rm diag}(\rho, \ -\rho, \ -\rho, \ 0).
\label{TDw} \ee
 One expects, that in the case of planar domain wall, the
 solutions to the full system of the nonlinear equations may in principle be obtained by
 analytic methods or by the numerical calculations.

Information on the phase transition, produced by the domain wall  of the
chiral field model, is concentrated in the structure of the stress-energy
tensor $T^{(mat)}_\mn ,$  (\ref{Tmat}) which should be matched with  phase transition
in gravitational sector in accord to Einstein equations.
Gravitational counterpart to this phase transition is the transfer
from the flat  metric inside the bubble, $r<r_e ,$ to the external metric of the exact
KN solution. For this transfer there is a remarkable ansatz suggested by G\"urses
and G\"ursey (GG) \cite{GG}. The standard  KS form of the metric (\ref{ksH}) with the
fixed Kerr congruence $k^\m ,$ is deformed only in the form of
function $H $ (\ref{H}), which takes the form   \be H=f(r)/\Sigma, \quad \Sigma=(r^2 +
a^2 \cos ^2\theta) \label{HGG} ,\ee where $f$ is arbitrary smooth function.

For the external KN metric $r>r_e ,$ one sets $f(r)=f_{KN}= mr -e^2/2 ,$ while the
smooth transfer to some internal metric may be provided by any its smooth
extension $f(r)=f_{int}.$ The GG form of metric has remarkable properties.
 It was shown in \cite{BurBag,BEHM} that setting for the interior
 $f(r)=f_{int}=\alpha r^4,$ one obtains a regular version of the KN
 metric, in which the Kerr singularity is suppressed.
 The GG form of metric describes the
 rotating solutions with flat asymptotic as well as the rotating
versions of the de Sitter and Anti-de Sitter solutions. Moreover,
it allows one to match smoothly  the external and internal metrics
of different sorts \cite{BurBag,BEHM}. On the
other hand the GG form of metrics is a particular case of the
Kerr-Schild solutions and inherits its  remarkable properties. In
particular, the electromagnetic field and its stress-energy tensor
display a partial linearization in the Kerr-Schild and
GG-spacetimes, and there is the exact correspondence between the
rotating and non-rotating solutions which allows one to simplify
analysis of the rotating solutions, by means of the analysis of
its non-rotating analogs. In particular, the
stress-energy tensor of the KN solution and its non-rotating
analog with $a=0$ take in the orthonormal tetrad $u, l, n, m ,$
where $ u$ is the unit timelike vector and $l$ the radial one, the
 form \cite{BurBag,BEHM,GG,Dym}
\begin{equation}
T_{\mn} = (8\pi)^{-1} [(D+2G) g_{\mn} - (D+4G) (l_\m l_n -  u_\m
u_\n)], \label{Tt}
\end{equation}
where
\begin{equation}
G= \frac{f'r-f}{\Sigma^2}, \quad D= - \frac{f^{\prime\prime}}
{\Sigma} \ , \label{GD}
\end{equation}
which may be  recognized as diagonal one
 \be T^\m_\n ={\rm diag}(\rho, \ -\tau, \ -\tau, \ p),
\label{TGG1} \ee where the radial pressure $p$ and tangential
stress $\tau$ are given by
\begin{equation}
\rho = - p = \frac{1}{8\pi} 2G,  \quad \tau = - \frac{1}{8\pi}(D+
2G)= - \rho - \frac{D}{8\pi}. \label{prtt}
\end{equation}

 This correspondence allows us to consider the phase transfer in the non-rotating
 case and translate the results to  the rotating KN
 solution.

For the case of quick rotation, $a>>r_e ,$ there appears a stringy
contribution to the mass-energy caused by concentration of the
electromagnetic field on the edge of bubble.
However, analytic calculations showed (see \cite{BurSol}) that the
`stringy' contributions from the shell and external em field are
mutually cancelled, and the total mass $m_{ADM}^{(total)} =
m_{ADM}^{(int)} + m_{ADM}^{(shell)} + m_{ADM}^{(ext)}$ turns out
to be equal to $m .$ Indeed, this result could be predicted a
priori, since the total ADM mass is determined {\it only} by the
asymptotical gravitational field, i.e. only by the value of parameter $m$ in
function $H ,$ (\ref{H}). Therefore, the naive stringy interpretation
 does not go, at least for the GG form of the metric.

The typical local structure of the stress-energy  tensor for the
vacuum domain wall \cite{IpsSik,CvGrifSol} is  \be T^\m_\n = \rho
\ {\rm diag} ( 1 , \ - 1 , \ - 1 , \ 0) \label{TplanDW} ,\ee
which shows that the surface energy density $\rho$ is equal to
tangential stress $\tau$.

Comparison between (\ref{TplanDW}) and (\ref{TGG1})
shows that  the GG metric contains the domain wall contribution,
 however there is no full correspondence.
 It opens a chance to get the the correspondence and stringy effect from some
generalizations of the GG form of metric. In particular, the
appearance of the string tension may be related with  extra
dilaton or axion  fields, which are typical for the metrics of domain wall
models and their analogs in superstring theory \cite{Cvet91}.
One more very important generalization of this problem may be related
with transfer from the used Einstein-Maxwell-Higgs (Chiral) system of the
eqs. to its remarkable natural analog in 4D Supergravity \cite{WesBag},
where the complex chiral fields $\phi^{(i)}$ form an extra K\"ahler manifold
supplied by K\"ahler metric and K\"ahler potential. It is probably, that the
solutions turn out to be consistent with such a generalized system.

We have to stop at this point, since nether  analytic calculations for planar
domain walls, nor the numerical calculations even for the  Einstein-Maxwell-Higgs
system were so far performed. There is a great field for activity.

\section{How the Dirac equation is hidden inside solitonic source of the KN electron.}

As we have shown, the  Kerr-Newman solution has many remarkable
properties indicating its relationships with the structure of the
Dirac electron. However, all these evidences cannot draw us away
from the natural question how and where the Dirac equation may be
residing in the  solitonic source of KN geometry. In this chapter
we will try to answer this question. The soliton source of the KN
represents a bag confining the Higgs field. In some respects this
bag is similar to the MIT and SLAC bag models.

Analyzing the twosheeted
structure of the KN solution, we have seen that the Kerr geometry
is based on the twistorial structures of the Kerr congruence. The
naked Kerr singular ring forms a branch line of space
 into the sheets of advanced and retarded
fields, and the  null vector field, $ k_\m (x)$ generates the
Principal Null Congruence (PNC) $\cal K ,$ form of which is changed by the
transfer from positive ($r>0$) to negative ($r<0$) sheet. The
surface $r=0$ represented a disklike ``door" separating two
different null congruences ${\cal K}^\pm ,$ creating two different
metrics \be g_\mn^\pm =\eta_\mn + 2H k_\m^\pm k_\n^\pm
\label{KSpm} \ee on the same Minkowski background $M^4.$

 It seemed that the formation of the soliton source closed this ``door" and removed
 the problem of twosheeted space. However, this problem emerges from
 another side. The second sheet appeared as a sheet of advanced fields, which are
 related with the old Dirac problem of  radiation reaction
 \cite{DirRad}. Dirac splits the expression for retarded potential $A_{ret}$
 into a half-sum and half-difference with advanced fields $A_{adv}$ as follows \be
A_{ret} = \frac 12 [A_{ret} + A_{adv}] + \frac 12 [A_{ret} -
A_{adv}].\label{DirSplit} \ee The half-sum he takes responsible for
 self-interaction of the source, while the half-difference should be
responsible for radiation and radiation reaction. The soliton source
acquires the second sheet outside the bag as the sheet
 of advanced fields, which should provide self-interaction
 of the soliton and create a massive Dirac equation.
 In accord with the basic properties of the Kerr-Schild solutions,
 the fields $A_{ret}$ and $A_{adv}$ could not reside on the same physical sheet,
because each of them should be aligned with the its own  Kerr
congruence. Considering the retarded sheet as a basic physical
sheet, we fix the congruence ${\cal K}_{ret} $ and corresponding
 metric $g_\mn^+$, which are not allowed for the advanced field $A_{adv}.$
  The advanced field is to be consistent with another congruence ${\cal K}_{adv} $,
  which should be positioned on a separate sheet with different metric $g_\mn^-$,.
 However, the problem of their incompatibility disappears inside the bag,
 where the space is flat, and the both null congruences ${\cal K}_{ret} $ and ${\cal
K}_{adv} $ are null not only with respect to the corresponding
Kerr-Schild metrics, but also with respect to the flat  Minkowski
background.

\begin{figure}[ht]
\centerline{\epsfig{figure=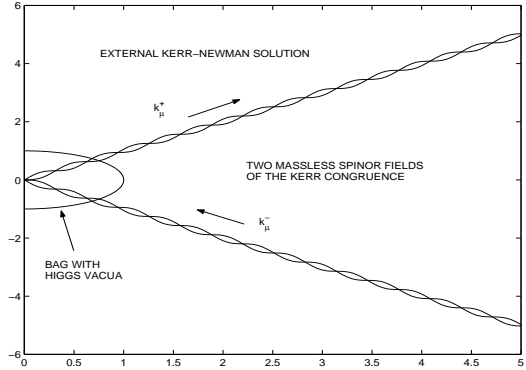,height=5cm,width=7cm}}
\caption{Creation of the Dirac equation inside the bag by the
confined Higgs field coupled to two  massless spinor fields
$k_\mu^+$ and $k_\mu^-$, generators of the Kerr principal null
congruences $\cal{K}^+$ and $\cal{K}^-$.}
\end{figure}

 For the sake of convenience, we replace further notations
 ${\cal K}_{ret} $ and ${\cal K}_{adv} $
 by the notations ${\cal K}^\pm ,$ identifying ${\cal K}^+ = {\cal K}_{ret}, \
 {\cal K}^- = {\cal K}_{adv} .$  The external null fields
$k_\m^\pm (x)$ are extended  inside the bag \emph{on the same
sheet} forming two different null congruences determined by two
conjugate solutions of \emph{the Kerr theorem} $Y^\pm (x^\m),$
\cite{DKS,KraSte}.

Now, we have to say a few words on the \textbf{Kerr theorem}. The
Kerr theorem is formulated in the Minkowski background,
$x^\m=(t,x,y,z) \in M^4 ,$ in  terms of the projective twistor
coordinates \be T^A= \{ Y, \ \z - Y v, \ u + Y \Z \},  \qquad A=1,2,3 ,
\label{(TA)} \ee where $u,v,\z,\Z$ are the null Cartesian
coordinates, $ \z = (x+iy)/\sqrt 2 ,\quad  \Z = (x-iy)/\sqrt 2 , u
= (z + t)/\sqrt 2 ,\quad v = (z - t)/\sqrt 2.$ The Kerr congruence
is determined by solution of the equation \be F(T^A) =0
,\label{FTA}\ee
 where $F$ in general case is a holomorphic function on the projective twistor space $CP^3.$
For the Kerr solution the congruence is created by function $F$
which is quadratic in $Y$ and may be represented in the form
 \be F(Y,x^\m) = A(x^\m) Y^2 +
B(x^\m) Y + C(x^\m), \label{FKN} .\ee In this case the equation
(\ref{FTA}) has two explicit solutions \be Y^\pm (x^\m)= (- B \mp
\tilde r )/2A, \label{Ypm}\ee where $\tilde r= (B^2 - 4AC)^{1/2}
,$ and it was shown in \cite{BurTMP}, that these two solutions are
related by antipodal correspondence \be Y^+ = - 1 /{\bar Y^-}
\label{antipY}. \ee

It should be noted, that $Y$ is a projective spinor coordinate, $Y
=\phi_1/\phi_0,$ and it is equivalent to the Weyl two-component
spinor  $\phi_\alpha =(\phi_1,\phi_0$)$^T$.

 In the Kerr-Schild formalism \cite{DKS}, function $Y(x^\m)$ determines
 the null congruence as a field of null directions $k_\m (x^\m)$ via differential form
 \be  k_\m dx^\m
=P^{-1} (du + \bar Y d \zeta + Y d \bar\zeta - Y\bar Y dv) ,
\label{e3} \ee where $ P=(1+Y\Y)/\sqrt 2$ is a normalizing factor.
This form is equivalent to standard representation of the null
vector via spinor $\phi$ and Pauli matrices $k_\m^+ =\bar \phi
\bar \sigma_\m \phi ,$ and therefore, generating two conjugate
null congruences, the Kerr theorem determines simultaneously two
massless spinor fields of different chirality $\phi_\alpha =(\phi_1,\phi_0$)$^T$
and $\chi^{\dot\alpha} =
(\chi^{\dot 1},\chi^{\dot 0})^T$

  We can compare these spinor fields with the structure of the Dirac equation
  \be (\gamma^\m \hat \Pi _\m
+m)\Psi=0, \label{InitD} \ee in which $\Psi =
(\phi _\alpha, \chi ^{\dot \alpha})^T$
and $ \hat \Pi _\m = - i  \d _\m - e
A_\m $, and $ \hat \Pi _\m = - i  \d _\m - e A_\m $, written in
the Weyl basis, where it  splits  in two equations \cite{BLP} \be
 \bar\sigma ^{\m \dot\alpha \alpha} (i \d_\m  +e A_\m)
 \phi _{\alpha} =  m \chi ^{\dot \alpha}, \quad  \sigma ^\m _{\alpha \dot \alpha} (i \d_\m  +e A_\m)
 \chi ^{\dot \alpha}=  m \phi _\alpha .
\label{Dir} \ee In the Standard Model these equations are called
the ``left handed" and the ``right handed electron fields". In the
massless case (\ref{Dir}) represents two equations, for two Weyl spinors,
connected by antipodal relation (\ref{antipY}). The corresponding null vectors
\be k^\m_{L} = \bar\phi \sigma^\m \phi \ , \quad k^\m_{R} =
\bar\chi \bar\sigma^\m \chi, \quad  k_{\m L} k^\m_L =\ k_{\m R}
k^\m_R =0 ,\label{kLR} \ee describe two principal null congruences
given by antipodal solutions $Y^\pm$ of the Kerr theorem. Outside
the bag these null fields should reside on different sheets of the KN
solution, but penetrating inside the bag they are meeting without
conflict and can reside on the same false-vacuum sheet because of flatness of
the space inside the bag. bag. Outside the bag they form a four-component Dirac spinor
$\Psi (x)$ satisfying the massless Dirac equation
\be (\gamma^\m
\hat \Pi _\m )\Psi=0, \label{InitD0} \ee
which acquires mass term inside the bag from the Higgs field $\Phi $ through the Yukawa interaction
 \be
{\cal L}_{Yukawa}(\Phi,\Psi) = - g \bar\Psi \Phi \Psi \label{Yuk}.
\ee Therefore, two antipodal twistorial congruences, obtained as
two conjugate solutions of the Kerr theorem $Y^\pm (x)$, create
two massless spinor fields $\phi _\alpha$ and  $\chi ^{\dot
\alpha}$, which are combined in a four-component massless Dirac
field  $\Psi = (\phi _\alpha, \chi ^{\dot \alpha})^T$.
The  Higgs field $\Phi$ confined inside the bag connects them by the
Yukawa coupling (\ref{Yuk}), which gives the mass term to the
Dirac equation in full agreement with the one of principal tasks
of the Higgs field in the Standard Model.

\section{Outlook: relation to superstring theory.}

\textbf{Solitonic regularization}
The negative sheet of the
metric disappears and metric turns out to be regularized and
practically flat. The model contains the lightlike heterotic
string on the border of the bubble, however, it is axially
symmetric and the traveling waves are absent. Therefore, some
extra excitations are needed to create the traveling
waves.

Does the KN model of electron contradict to Quantum Theory? It
seems ``yes'', if one speaks on the "bare" electron. However, in
accordance with QED, vacuum polarization creates in the Compton
region a cloud of virtual particles forming a "dressed" electron.
This region gives contribution to electron spin, and performs a
procedure of renormalization, which determines physical values of
the electron charge and mass. We describe here an alternative
solitonic version of the regularization performed on the base of
the Higgs field, which seems to us a more physical model contrary
to the used in QED  formal mathematical model.

Speaking on the ``dressed'' electron, one can say that the physical
contradiction between the KN model and the Quantum electron in QED
is absent. Note however, that dynamics of the virtual particles in QED is
chaotic and can be conventionally separated from the
``bare''electron. In the same time, the vacuum state inside the
Kerr-Newman soliton forms a {\it coherent oscillating state} joined
with a closed Kerr string. The oscillating bubble source represents
an {\it integral whole with the extended electron,} and its `internal'
structure cannot be separated from a ``bare'' particle.

 The
KN electromagnetic field is regularized since the maximal value of
the vector potential  is realized in the equatorial plane, on the
stringy boundary of the bubble.
 Note, that radius
of the regularized closed string, being shifted to the boundary of
the bubble, turns out to be slightly increased. The position of
the string confirms the known suggestion by Witten that the heterotic
strings have to be formed on the boundary of a domain wall \cite{WitAxi}.

It should be noted that the real closed Kerr string is only the peak of the iceberg.
As it was shown in the recent paper \cite{BurAlter} there is deep parallelism between the
complex structure of the Kerr geometry and basic structures
of the superstring/M-theory. In particular, about two decades ago, author obtained a
complex string inside of the complex 4D Kerr geometry, which together with the closed
 heterotic string of the KN source  forms a membrane source of the
M-theory. The closed heterotic Kerr string is light-like, and its structure is similar to the Sen
fundamental string solution to low-energy heterotic string theory.
In the recent paper \cite{AdNew} Adamo and Newman reobtained these two strings analyzing
asymptotic form of the geodesic and shear-free congruences. Their emotionally comments
are worth quoting:  ``...It would have been a cruel god to have layed down such a pretty scheme
and not have it mean something deep.''

In the recent papers \cite{BurAlter,BurTMP}, there was obtained a new remarkable fact --
appearance of the Calabi-Yau twofold (K3 surface) in the  projective twistor space $CP^3 .$
Therefore, the famous K3 surface of the superstring theory represents nothing other than a
twistorial description of the Principal Null Congruence (PNC) of the Kerr geometry.
One can suppose, that this parallelism is not accidental, and there should be an underlying
superstring structure lying beyond these relationships, and we suggest that it is N=2
critical superstring.
The N=2 superstring is one of the three string theories consistent with quantum
 theory; number of supersymmetries N= 0, 1, 2 corresponds to
 consistent space-time dimensions D=26, D=10 and D=4, \cite{GSW}.
 The N=2 string, having complex dimension two, was very popular a few
decades ago \cite{GSW}. However, it has unusual signature which is
conflicted with the real minkowskian space-time. As a result, it
was almost forgotten, and in particular it does not discussed in
the very good modern textbook \cite{BBS}. It has been shown
recently in \cite{BurAlter}, that the N=2 string can be embedded in
the complex Kerr geometry, and moreover, there are evidences that
it can be considered as a complex source of the Kerr geometry.

\section*{Acknowledgments}
This work is supported under the RFBR grant 13-01-00602. Author thanks Theo M. Nieuwenhuizen
for permanent interest to this work and useful conversations.

\section{References}


\end{document}